# Title: Ultrathin Terahertz Planar Lenses

**Authors:** Dan Hu[1,2,†], Xinke Wang[1,†], Shengfei Feng[1], Jiasheng Ye[1], Wenfeng Sun[1], Qiang Kan[3], Peter J. Klar[4], and Yan Zhang[1,2,*]

**Affiliations:**

[1]Department of Physics, Capital Normal University, Beijing Key Lab for Terahertz Spectroscopy and Imaging, and Key Laboratory of Terahertz Optoelectronics, Ministry of Education, Beijing 100048, China

[2]Department of Physics, Harbin Institute of Technology, Harbin 150001, China

[3]State Key Laboratory for Integrated Optoelectronics, Institute of Semiconductors, Chinese Academy of Sciences, Beijing 100083, China

[4]Institute of Experimental Physics I, Justus-Liebig University, Heinrich-Buff-Ring 16 35392 Giessen, Germany

*Correspondence to: Email: yzhang@mail.cnu.edu.cn

†These two authors contributed equally to this work.

**Abstract**: Conventional optical components shape the wavefront of propagating light by adjusting the optical path length, which requires the use of rather thick lenses, especially for the adjustment of terahertz (THz) radiation due to its long wavelength. Two ultrathin THz planar lenses were designed and fabricated based on interface phase modulation of antenna resonance. The lens thicknesses were extremely reduced to 100 nm, which is only 1/4000$^{th}$ of the illuminating light wavelength. The focusing and imaging functions of the lenses were experimentally demonstrated. The ultrathin optical components described herein are a significant step toward the development of a micro-integrated THz system.

**One-Sentence Summary:** Ultrathin planar lenses with a thickness of 100 nm were designed and fabricated to implement THz beam focusing and imaging.

Terahertz (THz) radiation lies in the frequency range between infrared and microwaves, typically having wavelengths ranging from 10 μm to 3 mm. THz technology is developing rapidly in many independent fields and has many potential applications *(1)*. However, due to the relatively long wavelengths of THz radiation, most THz components, such as lenses and prisms, are on a large scale and are not suitable for system integration.

Conventional optical components shape the wavefront of propagating light via gradual phase changes that accumulate along the optical path, usually via alterations to the spatial distribution using the thickness or refraction index of the components. Early optical components possessed continuous curved surfaces to achieve phase modulation, as indicated in Fig. 1A. This continuity determines the bulkiness of the components. Further technological developments utilized the 2π phase jump to reduce component thickness to the wavelength scale, as shown in Fig. 1B. Subsequently, metamaterials with extremely large effective refractive indices have been used to further reduce the thickness of the optical components *(2, 3)*. However, the basic theory for wavefront shaping is still based on phase accumulation along the optical path, and the thickness of the corresponding components is still quite large. The question remains as to whether it is possible to further reduce the thickness of the optical components.

Alternatively, phase changes can also be introduced by an optical resonator. Electromagnetic cavities *(4-6)*, nanoparticles clusters *(7, 8)*, and plasmonic antennas *(9,10)* have previously been employed for tailoring phase changes. Recently, a novel method was proposed to introduce a phase discontinuity at the interface between two media *(11-13)*. In this method, the geometry of planar V-shaped antennas was spatially selected, and the phase shift between the emitted and illuminating lights could be controlled arbitrarily; the generalized laws of reflection and refraction using this method were described. If the antennas are spatially arranged according to a customized phase distribution, an ultrathin planar optical component can be utilized, as shown in Fig. 1C. In this case, the thickness of the component is far less than the incident wavelength.

We designed and fabricated complementary V-shaped antenna arrays in a 100 nm thick gold film to introduce abrupt phase shifts for focusing light with a 400 μm wavelength and imaging objects. The phase shifts were spatially arranged according to the wrapped phase distributions of cylindrical and spherical lenses. This THz focal plane imaging technology was utilized to characterize the properties of the fabricated samples. A sub-wavelength focal spot and clear images were achieved.

Eight basic complementary V-shaped antennas were selected to provide the desired phase changes (from 0 to $2\pi$ with $\pi/4$ intervals) and equal intensity modulation for 400 μm wavelength cross polarized light *(11)*. The size of a unit cell was 200 μm×200 μm. The total number of cells was 40×40; thus, the size of the lens was 8 mm×8 mm. The antennas were arranged according to the wrapped phase distribution of a cylindrical lens with a focal length of 4 mm, as shown in Fig. S1B. The 100 nm thick antenna arrays were fabricated on a 500 μm thick silicon substrate using standard optical lithography technology. Part of the cylindrical lens is shown in Fig. 2A. The phase shift was varied along the *x* direction and remained unvaried along the *y* direction *(14)*. The performance of the designed cylindrical lens was theoretically checked using the finite difference time domain method *(15)*. The intensity distribution of the cross polarized light in the *x-z* plane is shown in Fig. 2B. The THz wave was focused well on the preset position, which was 4 mm away from the lens.

This THz focal plane imaging system was employed to measure the complex field distributions around the focal plane of the fabricated cylindrical lens *(14, 16, 17)*. The scanning step along the *z* direction was 0.25 mm. The intensity distribution of the focused THz beam around the focal plane is presented in Fig. 2C, which is consistent with the simulation result shown in Fig. 2B. The intensity distributions along the white dashed lines shown in Figs. 2B and 2C are plotted in Fig. 2D, which exhibit good Gaussian profiles with a full width at half maximum (FWHM) of 270 μm. In comparison, according to classical optics, the focal spot size of the lens is given by $8f/(kD) = 255$ μm, where $f$ is the focal length of the lens, $k$ is the wave vector of the incident light, and $D$ is the diameter of the lens *(18)*. The difference was likely caused by the quantization of the phase distribution and fabrication errors. As shown in Fig. 2E, a clear line focus on the preset plane was achieved as expected.

To demonstrate the imaging performance of the ultrathin planar lens, a spherical lens with a 4 mm focal length was also designed and fabricated, part of which is shown in the photograph in Fig. 3A.The intensity distribution of the cross polarized light on the preset focal plane is shown in Fig. 3B when the lens is illuminated with a linearly polarized THz wave. A satisfactory focal spot was observed. Three letter patterns drilled on a stainless steel slice, as shown in Fig. 3C, were used as imaging objects. Based on the dimensions of the THz beam and the detection crystal, the three letters C, N, and U were imaged separately. Their THz images are clearly displayed in Figs. 3D-3F, indicating that the ultrathin planar spherical lens performs well.

In summary, cylindrical and spherical ultrathin planar lenses were designed for THz focusing and imaging. The thicknesses of the lenses were only $1/4000^{th}$ of the illuminating wavelength, which is quite promising for their use in system integration. Lens thickness can be further reduced because the skin-depth of metal in the THz range is only several tens of nanometers *(19)*. This method can also be extended to design polarization converters, wave plates, and optical interconnection devices.

**Acknowledgments:** This work was supported by the 973 Program of China (No. 2011CB301801), the National Natural Science Foundation of China (No.10904099, 50971094, and 11174211), and the Beijing Natural Science Foundation (Grant No. KZ201110028035).

**Fig. 1**. Three mechanisms for lens design. **(A)** A conventional lens that shapes the wavefront of light via phase changes accumulated along the optical path. **(B)** A Fresnel lens that removes the surplus thickness corresponding to multiple $2\pi$ phase changes. **(C)** An ultrathin lens that localizes the phase change at the interface between two media via antenna resonances.

**Fig. 2. (A)** Photograph of a part of the fabricated cylindrical lens. The size of each cell was 200 μm×200 μm, and the silt width was 5 μm. The thickness of the complementary V-shaped antenna was 100 nm; the antenna was fabricated on a 500 μm thick silicon substrate. The focal length of the lens was 4 mm at a wavelength of 400 μm. **(B)** Intensity distribution of the cross polarized light for the designed cylindrical lens along the *z* direction as simulated using commercial software according to the FDTD method. **(C)** Experimental measurement of the intensity distribution of the cross polarized light for the fabricated cylindrical lens on the *x-z* plane. The step along the optical propagation direction was 0.25 mm. **(D)** Intensity distributions along the white dashed lines shown in (B) and (C). The blue solid and red dotted curves are the simulated and experimental results, respectively. The full width at half maximum is 270 μm, which approaches the resolution of a conventional lens with the same numerical aperture. **(E)** The line focus of the cylindrical lens on the preset focal plane in experiments.

**Fig. 3. (A)** Photograph of the center part of the fabricated spherical lens. Antennas were arranged according to the phase distribution of a spherical lens with a 4 mm focal length. **(B)** The focal spot of the spherical lens on the preset focal plane achieved in experiments. **(C)** The object to be imaged. The size of each letter was 4 mm×5 mm, and the slit width was 1 mm. The letters were drilled on a stainless steel slice with a thickness of 0.3 mm. **(D)-(E)** Images of the three letters. The images were reduced according to the object-image relationship.

**Supplementary Materials:**

Materials and Methods

Figures S1 and S2

References (*19-20*)

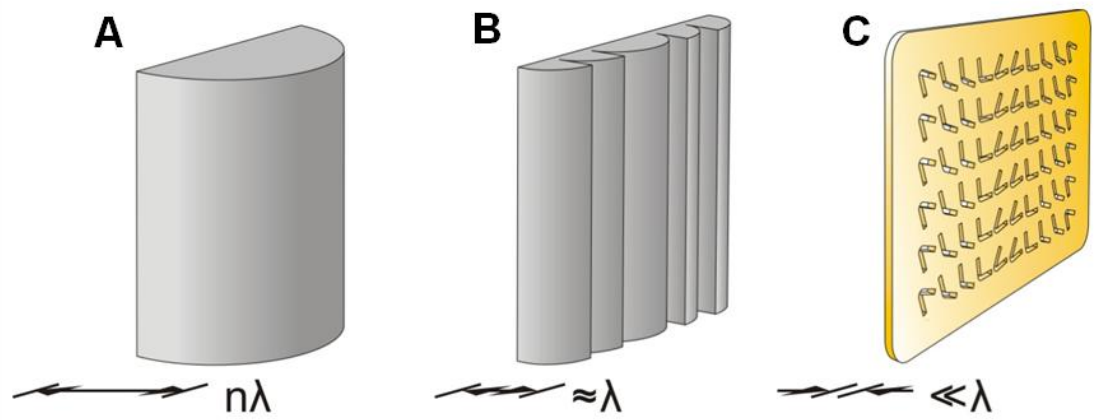

Fig. 1

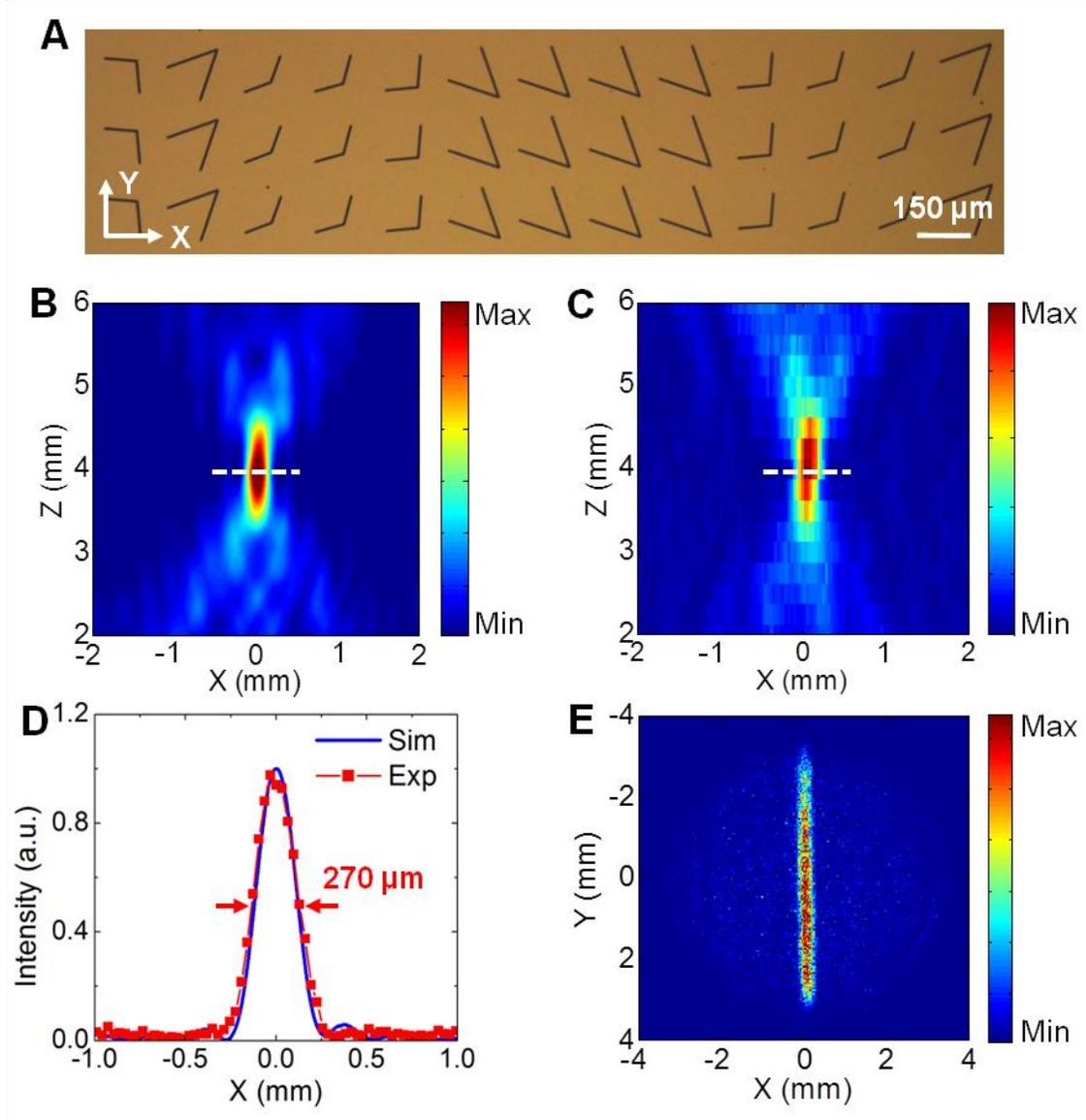

Fig. 2

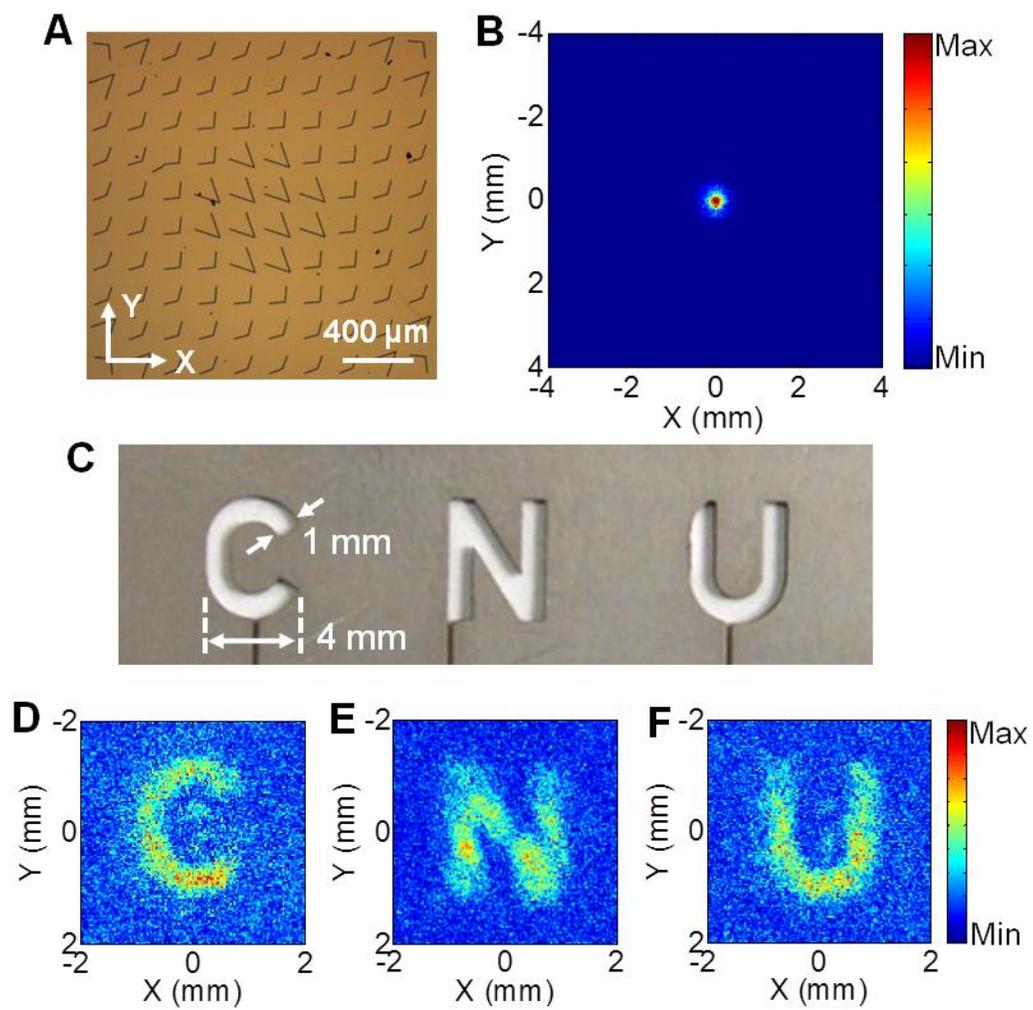

Fig. 3

## Supplementary Materials:

### Design method

For this design, the selected, complementary V-shaped antenna, which is displayed schematically in the inset of Fig. S1A, was composed of two equivalent rectangular slits connected at one end with an angle $\theta$ *(19)*. The width and length of the rectangular slit are *w* and *h*, respectively. The frequency of the illuminating light is 0.75 THz, and the corresponding wavelength is 400 μm. In this design, the width of the rectangular slit (*w*=5 μm) and the angle between the angle bisector line of the V-shaped antenna and the *y* axis ($\beta$=45°) were fixed, and the size of each unit cell (200 μm×200 μm) was also fixed for all of the V-shaped antennas to avoid coupling between two adjacent antennas. Eight complementary V-shaped antennas were designated as the building blocks. The first four antennas had $\theta$=60°, 100°, 120°, and 130° with corresponding lengths of *h*=150, 90, 82, and 78 μm. Four other antennas were mirror images of the first four antennas, as shown in Fig. S1A. Because the antenna array was characterized with the pulse THz source, only radiation with the designated frequency (0.75 THz) will achieve resonance with the antennas and be able to transmit the sample with high efficiency; the metal film will block other frequency components. Therefore, devices constructed using this type of antenna array have two basic functions: wavefront shaping and filtering.

Numerical simulations based on the finite-difference time-domain (FDTD) method were employed to ensure that the design was accurate. The dimensions of the simulated structure were 200 μm×200 μm×2000 μm, which allowed a complementary V-shaped antenna to be placed on the *x-y* plane and sufficient space for scattered light monitoring. Periodic boundary conditions were applied in the *x* and y directions, and a perfectly matched layer (PML) boundary condition was applied in the *z* direction. A plane wave at 0.75 THz (400 μm wavelength) with the *x* polarization direction was used as the incident light, and the amplitudes and phases of the scattered cross-polarized radiation were monitored. The amplitudes and phases of the scattered cross-polarized field arising from the eight antennas are shown in Fig. S1A, which clearly shows that the scattered fields have the same amplitudes and a constant phase difference of $\pi/4$.
Based on the numerical simulation results, two ultrathin planar lenses, a cylindrical lens and a spherical lens were designed. The required phase distribution of the cylindrical lens at different positions for *x* can be readily obtained based on the equal optical path principle: $\varphi(x) = 2n\pi + \frac{2\pi}{\lambda}(f - \sqrt{f^2 + x^2})$, where *n* is an arbitrary integer, $\lambda$ is the incidence wavelength, and *f* is focal length; the phase value is thus wrapped into the range of 0 to $2\pi$ and quantized to eight values. A set of complementary V-shaped antennas was selected according to the phase shift, as shown in Fig. S1B. Each lens (8 mm×8 mm) had 40×40 cells. The lenses were fabricated as ultrathin metal films (100 nm thickness) and were deposited on a 500 μm thick, double-sided polished silicon substrate using conventional photolithography and metallization processes.

**Experiment setup and measurements**

A terahertz (THz) balanced electro-optic focal plane imaging system was used to characterize the performance of the fabricated cylindrical and spherical lenses. Fig. S2A shows the experimental setup. Ultrafast 100 fs, 800 nm laser pulses with a 1 kHz repetition rate and an average power of 500 mW were divided into the pump and probe beams to generate and detect the THz waves, respectively. A <110> ZnTe crystal (3 mm thick) was illuminated by the pump beam, which passed through a concave lens (L1) with a 50 mm focal length to radiate a divergent THz beam via the optical rectification effect. A parabolic mirror with a 150 mm focal length collimated the THz waves, and the diameter of the THz beam was approximately 24 mm. The collimated THz beam was incident onto the lens. A second <110> ZnTe crystal (3 mm thick) was used as the sensor crystal to detect the modulated THz beam. A half wave plate (HWP) and a polarizer were employed to adjust the polarization of the probe beam. A 50/50 non-polarization beam splitter (BS) then reflected the probe beam to impinge on the sensor crystal. In the experiment, the polarizations of the incident THz beam and the probe beam were horizontal and vertical, respectively. The <001> axis of the sensor crystal was parallel to the incident THz polarization for measurement of the vertical polarization component of the THz field transmitted by the sample. The polarization of the probe beam was modulated by the THz field in the sensor crystal via the Pockels effect, and the two-dimensional THz image information was transferred to the change of the probe polarization. The reflected probe beam was captured by the imaging component, which was composed of a quarter wave plate (QWP), a Wollaston prism (PBS), two convex lenses (L2 and L3), and a CY-DB1300A CCD camera (Chong Qing Chuang Yu Optoelectronics Technology Company). The QWP was used to impart a $\pi/2$ optical bias to the probe beam, and the PBS was used to split the probe beam into two mutually orthogonal, linearly polarized components. Their images were projected onto the CCD camera by the 4f system consisting of L2 and L3. The difference between the two images was extracted to analyze the THz information *(20)*. The CCD camera was synchronously controlled with a mechanical chopper to capture images with a 2 Hz frame ratio. Here, 100 frames were averaged to enhance the signal to noise ratio of the system. The acquired image was 300×300 pixels. Varying the optical path difference between the THz beam and the probe beam enabled the collection of THz images at each time-delay. The time window was 33 ps, and the time resolution was 0.13 ps. The THz temporal signal on each pixel was determined by the Fourier transform method, and the 0.75 THz component was extracted to build the THz image.

   A *z*-scan measurement was performed for the designed cylindrical lens with a focal length of 4 mm, as shown in Fig. S2B. The original distance between the lens and the sensor crystal was 4 mm. The scan range was from -2 mm to 2 mm before and behind the focal plane, and the scanning step was 0.25 mm. On each scanning plane, the 0.75 THz intensity distribution was acquired to compose the cross-sectional image of the focal THz beam along the *z* axis, as shown in Fig. 2C.

   An imaging experiment was performed for the designed spherical lens with a focal length of 4 mm, as shown in Fig. S2C. In the experiment, the objects had hollowed patterns of the letters "C", "N", and "U" on a stainless steel slice with a thickness of 0.3 mm, as shown in Fig. 3C. The size of each letter was 4 mm×5 mm. The distance between the objects and the lens was 12 mm, and the distance between the lens and the sensor crystal was 6 mm. The reduced real images of the objects were projected onto the sensor crystal, and their 0.75 THz images are presented in Figs. 3D-3F.

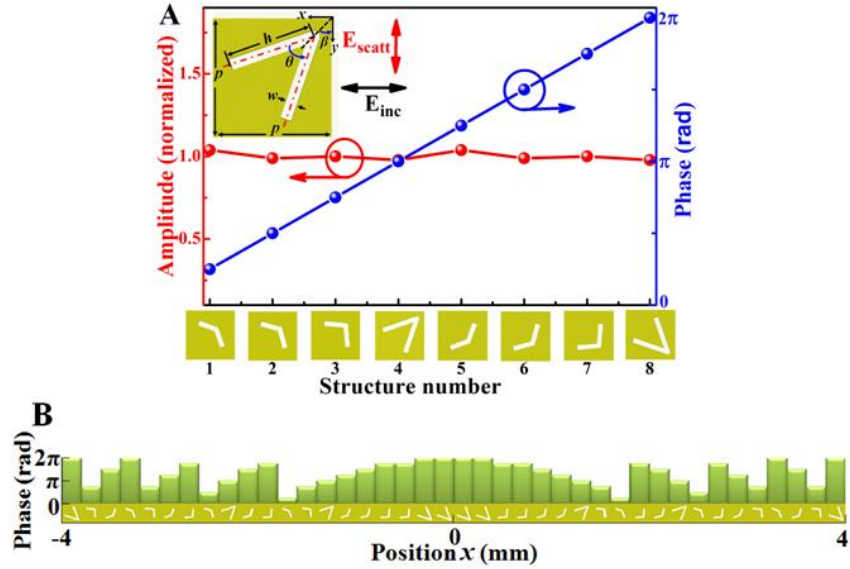

**Fig. S1.** (**A**) Simulated scattering amplitudes and phases of the cross-polarized radiation from individual complementary V-shaped antennas in a unit cell on a 500-μm-thick silicon substrate. Images of the selected eight inverse V-shaped antennas correspond to different phase delays. The inset shows a schematic view of a complementary V-shaped antenna. (**B**) Phase shifts at different *x* positions of the cylindrical lens (4 mm focal length); the corresponding complementary V-shaped antennas are also depicted in the figure.

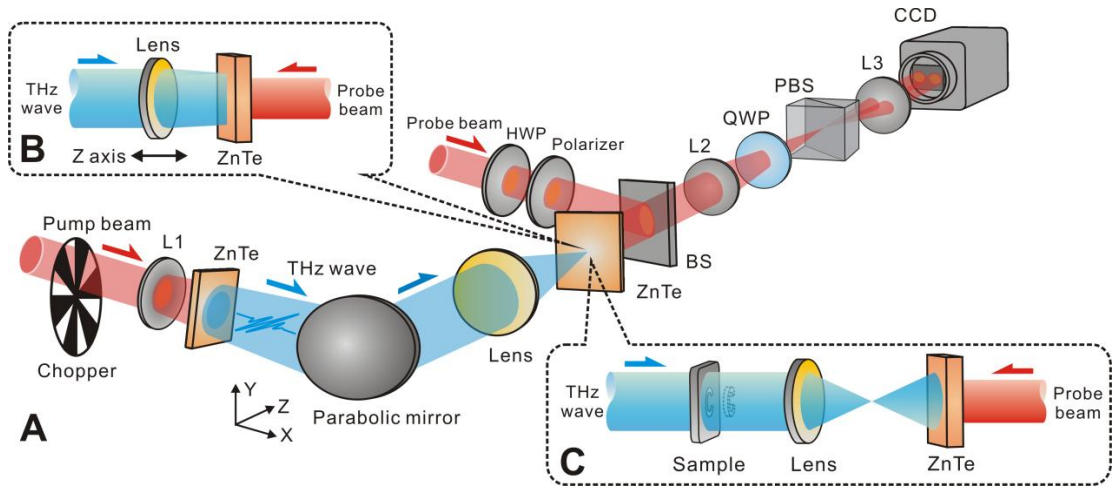

**Fig. S2.** (**A**) Terahertz balanced electro-optic focal plane imaging system. (**B**) The *z*-scan measurement of the ultrathin planar cylindrical lens (4 mm focal length). (**C**) The imaging experiment based on the ultrathin planar spherical lens (4 mm focal length).